# A Resonant-Tunneling Quantum-Dot Demultiplexer for Scalable Cryogenic Single-Electron Routing

Rajat Bharadwaj[1], Harshita Bose[2], and Madhu Thalakulam[1*]

[1]Indian Institute of Science Education & Research Thiruvananthapuram, Kerala 695551, India

[2]Delhi Technological University, New Delhi 110042, India

## Abstract

Scaling quantum electronic systems, such as quantum-dot spin qubits or single-electron devices, is increasingly limited by wiring and interconnect overhead between cryogenic devices and room-temperature control electronics. Analogous to demultiplexers and routers in classical communication networks, multi-channel quantum-dot architectures that selectively route electrons with a small number of shared control lines offer a promising route. Here, we propose and theoretically investigate a multi-channel, gate-defined quantum-dot architecture consisting of three parallel quantum dots that share a common source reservoir and couple to independent drain channels. Using an analytical electrostatic model and tight-binding quantum transport simulations, we show that our device, by exploiting resonant tunneling, can serve as an on-chip single-electron demultiplexer operating in the cryogenic regime. Closed-system eigenstate analysis confirms the presence of localized orbital states in each dot, while open-system transport calculations reveal resonant conductance peaks when these discrete energy levels align with the incident electron energy, in excellent agreement with the eigenenergy evolution obtained from the closed system. By engineering the barrier-gate spacing and the relative plunger-gate dimensions, we obtain well-separated resonances with minimal overlap between channels. Bond-current density calculations further show that transport is spatially confined to the resonant channel, with negligible leakage into neighboring outputs, yielding channel selectivity approaching unity across all resonant modes. The proposed architecture can, in principle, be adapted to established platforms such as Si/SiGe and GaAs/AlGaAs systems and also could potentially be extended to a larger number of channels, offering a route toward programmable quantum routing and future resonant-tunneling-based quantum electronic circuitry.

* madhu@iisertvm.ac.in

Gate-defined quantum dots formed on two-dimensional electron gas (2DEG) systems, owing to their scalability, discrete energy spectra, and electrostatic tunability, have attracted significant attention for applications in quantum electronics[1], single-electron devices[2–4], and more importantly, quantum-dot spin qubits[5–7]. While recent advances have substantially improved the scalability of quantum-dot qubits, the associated control infrastructure, including lines powering the gates, control electronics, and readout circuitry, has emerged as a major bottleneck to further scaling, particularly when these components must operate in close proximity to the quantum processor. The ability to switch electron flow between multiple transport channels is expected to play an important role in future cryogenic and quantum electronic systems. Analogous to demultiplexers and routers in classical communication networks, quantum transport structures capable of directing electrons toward selected output channels may enable programmable quantum interconnects, selective signal distribution, and reconfigurable nanoscale transport networks. Such architectures are especially relevant for large-scale low-temperature quantum systems, where the growing number of ohmic interconnects and control lines between room-temperature electronics and cryogenic devices presents a major scalability challenge[8,9]. Multi-channel electron routing architectures based on shared control and common transport reservoirs may therefore help reduce wiring complexity and improve scalability in future quantum electronic platforms. Existing reports have demonstrated electron routing using multiple phase-controlled gate signals to dynamically steer electrons[10,11]. However, in such approaches, the routed current is strongly dependent on the pulse frequency and is ultimately limited by the bandwidth of the gate lines[12,13]. An approach that leverages the same fabrication and gate-architecture principles used to realize the qubits themselves, while exploiting fundamental quantum transport phenomena for signal routing, would therefore be highly advantageous from both an architectural and operational standpoint.

Resonant tunnelling[14,15], a fundamental quantum phenomenon, enables highly selective and controllable electron transport, making such systems promising candidates for controllable nanoscale switching and routing applications. In this article, we propose and simulate a resonant-routing approach that uses multiple quantum dots, controlled by a set of common gates, to switch signals between different channels. The dots share a common barrier and plunger gates, and those are intentionally designed with different geometries for addressability. Owing to their different capacitive coupling strength, the dots exhibit distinct resonance peaks in transport, enabling selective current routing. Furthermore, the resonance-based architecture provides potential routes to control the transmitted current via source-drain biasing or dynamic modulation of the plunger voltage.

As a proof of concept, we model an architecture consisting of three gate-defined quantum dots sharing a common source reservoir and independent drain reservoirs. We calculate the electron wave-function using the Kwant package[16] and observe the formation of electron orbitals localized in each dot, confirming the presence of three well-defined parallel quantum dots. Using the closed-system eigenvalue spectrum, we evaluate transport characteristics at selected incident electron energies, revealing distinct resonant peaks for each quantum dot. The routing pathways are verified from the localized currents, the bond currents[17,18], which is obtained by the sum of individual contributions from the large number of lead modes to the dot state. Additionally, we quantify the channel selectivity and demonstrate nearly ideal routing performance. Our results demonstrate near-unit selectivity between different transport channels. The observed conduction peaks confirm that the transport happens via resonant tunneling. The bond-current analysis further reveals spatially selective current pathways associated with each resonant mode, and our simulation parameters agree very well with realistic device parameters. Together, these results establish the proposed architecture as a scalable platform for controllable multi-channel quantum transport, with promising applications in quantum routing, programmable electron networks, and quantum electronic circuitry.

The device under study consists of three parallel gate-defined quantum dots over a two-dimensional conducting channel, the two-dimensional electron gas (2DEG), as shown in Fig. 1(a). The dimensions of the device are provided in the Supplementary Material, S1. Within the tight-binding framework, the conducting channel is discretized on a square lattice with a computational lattice spacing $a$. The barrier and plunger gates are represented by brown and red colors, respectively. Gates of the same color are galvanically connected. The lower end of the 2DEG channel is connected to a common source lead, S, while the upper end is connected to three independent drain leads $D_1$, $D_2$ and $D_3$ separated by etch-defined regions. All surface gates are positioned 50 lattice sites above the 2DEG. The gate geometry is designed to create three spatially separated, charge-confined regions while maintaining independent transport paths to the corresponding drain leads. The central quantum dot is designed with a slightly larger barrier-gate separation (denoted L and L+ΔL in Fig. 1(a)) than the two side dots to minimize the effects of stray electrostatic coupling from the neighboring barrier gates. Furthermore, the plunger gates ($P_1$, $P_2$, $P_3$) are fabricated with different lateral dimensions to achieve distinct capacitive coupling strengths to the underlying electron gas and thereby facilitating selective electrostatic tuning of the three confinement regions. Similar to the barrier gates, plunger gates also share the same potential.

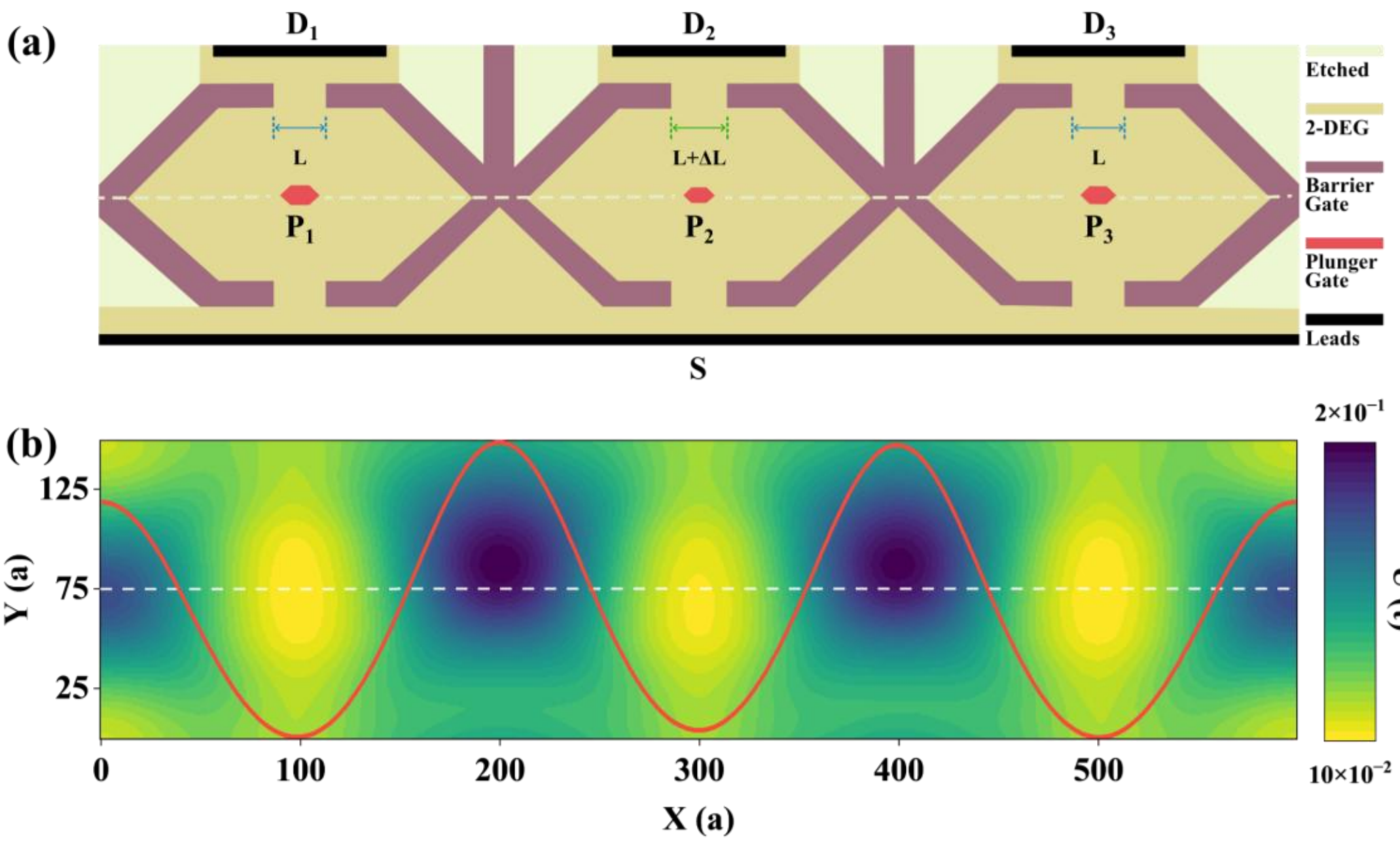

Fig. 1. (a) Device geometry of the proposed quantum-dot electron demultiplexer consisting of three parallel gate-defined quantum dots connected to a common source lead, S, and three independent drain leads $D_1$, $D_2$ and $D_3$. (b) Electrostatic potential energy profile of the 2DEG contributed by the gates for $V_b = -1t$ and $V_p = 0$, showing the formation of three spatially separated confinement regions. The red trace represents the potential at $Y = 75a$ , i.e., along the white dashed line shown in (a) and (b).

To determine the operational regime of the proposed device, we first calculate the potential energy of each lattice site using the analytical method developed by Davies *et al*[19]. For computational convenience, both the lattice constant $a$ and nearest-neighbor hopping energy $t$ are set to unity. $t$ is generally defined as $\hbar^2/2m_e^* a^2$ and can later be chosen specifically for the heterostructure. The Hamiltonian of the scattering region is then constructed as given in Equation (1).

$$(H_c)_{i,j} = \begin{cases} 4t + U_i, & i = j, \\ -t, & i, j \; nearest\; neighbors. \\ 0, & otherwise. \end{cases} \quad (1)$$

The term $U_i$ represents the onsite energy of the $i^{th}$ lattice site due to the applied gate potential $V_b = -1$ while $V_p = 0$, in units of the nearest-neighbor hopping energy $t$. The resulting energy profile, shown in Fig. 1(b), exhibits three distinct local minima beneath the plunger gates. The red trace in Fig. 1 (b) shows the line profile along $Y = 75$ (in units of $a$), represented by the white dashed line in Fig. 1 (a) and (b), reveals three well-separated potential wells. The matrix $H_c$ is then diagonalized to find eigenstates of the closed system.

For the prescribed gate voltages, the probability-density distributions corresponding to the six lowest-energy eigenstates, $E_{kl}$, are shown in Fig. 2. Although these eigenstates are obtained for the complete three-dot system, their probability densities exhibit strong spatial localization within the individual quantum dots, confirming the formation of well-defined confined orbitals[20–22]. Notably, the six lowest-energy eigenstates comprise two localized orbital states, the ground ($l = 0$) and first excited state ($l = 1$), for each quantum dot ($k \in \{1,2,3\}$, left to right dot), corresponding to the ground and first excited orbitals of that dot. $E_{30}$ and $E_{21}$ represent the lowest and highest eigenvalues, respectively. Thus, despite being eigenstates of the entire coupled system, the low-energy spectrum can be interpreted in terms of localized orbitals of the individual quantum dots.

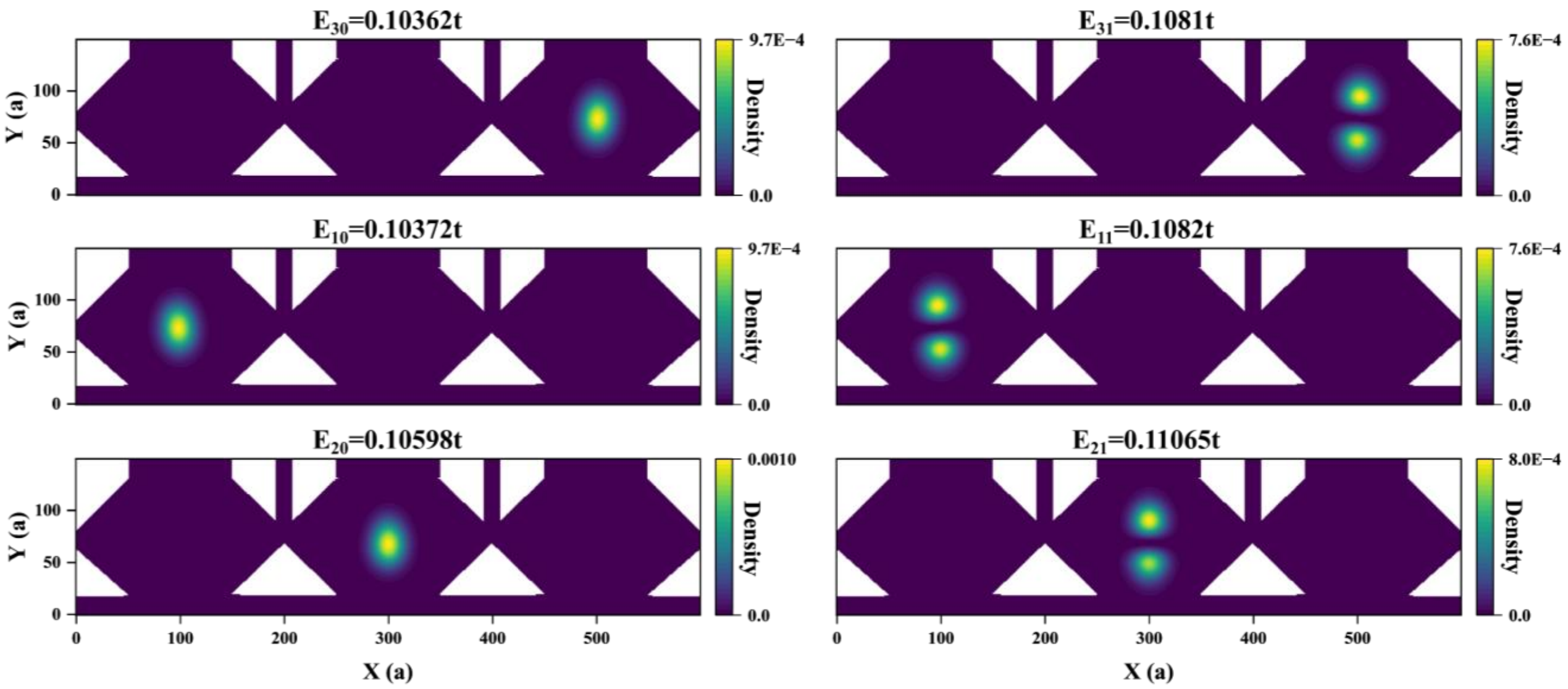

Fig. 2. Probability-density distributions of the six lowest-energy eigenstates of the closed system together with their corresponding eigen energies. The spatial localization of the wavefunctions confirms the formation of three distinct quantum dots that support discrete confined orbital states.

As the plunger-gate voltage is swept toward more negative values, the eigen-energies of the quantum-dot states, $E_{kl}$, evolve as shown in Fig. 3(a). Initially, the left and right-dot eigenstates are nearly degenerate, as expected from the approximately symmetric device geometry. As the gate voltage is reduced, however, the asymmetric electrostatic potential causes the two states to shift at different rates, increasing the energy separation between them.

Electron transport across the system is then investigated using the Kwant framework[16], which combines the tight-binding formalism with scattering theory to solve the time-independent Schrödinger equation and calculate the scattering matrix of the open system (with semi-infinite leads). Including the semi-infinite leads, the total Hamiltonian is written as Equation (2). $E_s$, represented by the horizontal dash-dot line in Fig. 3(a), is the energy of the incident electron from the source lead.

One can infer that sweeping the common plunger-gate voltage enables aligning the discrete dot levels with a high degree of selectivity between the three resonant tunneling channels.

$$H_{total} = \begin{pmatrix} H_{LD3} & 0 & 0 & 0 & \tau_{LD3} \\ 0 & H_{LD2} & 0 & 0 & \tau_{LD2} \\ 0 & 0 & H_{LD1} & 0 & \tau_{LD1} \\ 0 & 0 & 0 & H_{LS} & \tau_{LS} \\ \tau_{LD3}^{\dagger} & \tau_{LD2}^{\dagger} & \tau_{LD1}^{\dagger} & \tau_{LS}^{\dagger} & H_C \end{pmatrix} \quad (2)$$

Here, $H_{LS}$ and $H_{LD_i}$ denote the Hamiltonians of the source and drain leads[16], respectively, and $\tau_{LS}$ and $\tau_{LD_i}$ represent the coupling matrices between the central region and the corresponding source and drain leads. Hamiltonian of the leads follows a structure similar to (1), assuming $U_i = 0$ (i.e., the leads are assumed to be unaffected by the gate potentials). The transmission probability from the source lead $S$ to the drain lead $D_i$ for an electron with energy $E$ is then obtained from the scattering matrix of the system by solving the time-independent Schrödinger equation (3)[23,24], subject to open boundary conditions imposed by the semi-infinite leads.

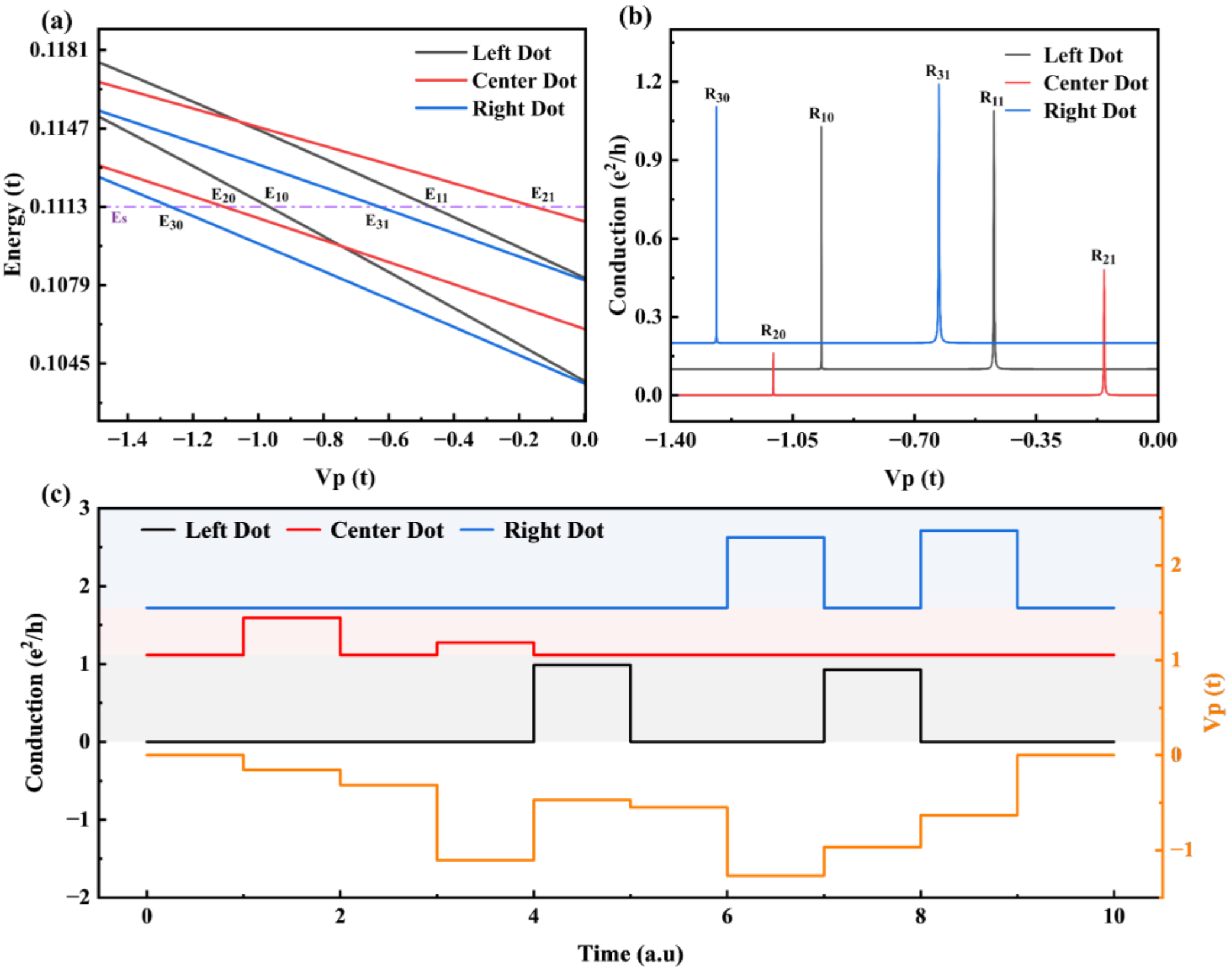

Fig. 3. (a) Closed-system eigenenergy spectrum as a function of plunger-gate voltage $V_p$. The dashed line indicates the selected electron energy $E_s$. Eigen energies of the whole system evolve unequally and intersect the $E_s$ at distinct points. (b) Conductance from the source lead to each drain lead, showing resonant peaks whenever a confined energy level aligns with the electron energy during transport. (c) Channel switching through common plunger-gate control, demonstrating selective electron routing among the three output channels. The applied plunger-gate voltage is shown in orange, while the conductance of the left, centre, and right dots are represented by the black, red, and blue traces, respectively. For visual clarity, the conductance traces are offset along the vertical axis.

$$H_{total}\Psi = E\Psi \qquad (3)$$

Here, $\Psi$ denotes the scattering-state wave function. The transmission coefficient[25,26] is then calculated as $T_{D_i,S}(E) = \sum_{m\in D_i}\sum_{n\in S}|S_{mn}(E)|^2$. $S_{mn}(E)$ is the scattering-matrix element representing the probability amplitude of transmission for an electron incident in a propagating mode $n$ of the source-lead to the propagating mode $m$ of the drain lead $D_i$. The corresponding conductance is obtained from the Landauer formalism[25] $G(E) = T(E)e^2/h$. We choose the incident electron energy as $E_s = 0.11129t$, slightly above the energy of the sixth eigenstate, $E_{21}$. As the common plunger-gate voltage is swept, conduction peaks $R_{kl}$ emerge in the conductance spectra, corresponding to resonant transport through the $l^{\text{th}}$ excited state from the source to the $k^{\text{th}}$ drain $D_k$, as shown in Fig. 3(b). The resonance peaks in conduction occur when the discrete dot energy levels align with the electron energy $E_s$ (denoted by the dashed line in Fig. 3(a)), demonstrating excellent agreement between the closed-system eigenenergy evolution and the open-system transport response. Since the chosen electron energy lies above only two confined states in each dot, two resonant peaks are observed for each channel. The spacing between successive resonances within the same dot reflects the corresponding orbital level spacing. The plunger gates are designed such that $P_2 < P_3 < P_1$ in geometry. The plunger gate $P_2$, being the smallest gate in the device, results in the largest apparent spacing between the central-dot conduction peaks compared with those of the left and right dots. Furthermore, the right plunger, for similar reasons, shows weaker capacitive coupling than $P_1$ and consequently a slower shift of the right-dot energy levels with $V_p$. This engineered asymmetry prevents overlap between neighboring resonances and enables clear separation of the conductance peaks in the prescribed operational regime. Additionally, the excited orbital states exhibit stronger coupling to the leads due to higher spatial overlap of the electron wave function in the dot with the lead modes[27], resulting in comparatively higher conductance values. This enables selective control of individual conduction channels, allowing the system to be switched between two distinct finite-conductance states by tuning the plunger-gate voltage. An example of this switching operation, implemented using a gate-voltage pulse sequence (orange), is shown in Fig. 3(c). The common plunger-gate voltage versus time is shown on the right axis, whereas the conductance through the different dots (left, center, and right) is shown on the left axis (black, red, and blue). The plot clearly shows well-defined switching capability between different channels using a common plunger-gate voltage.

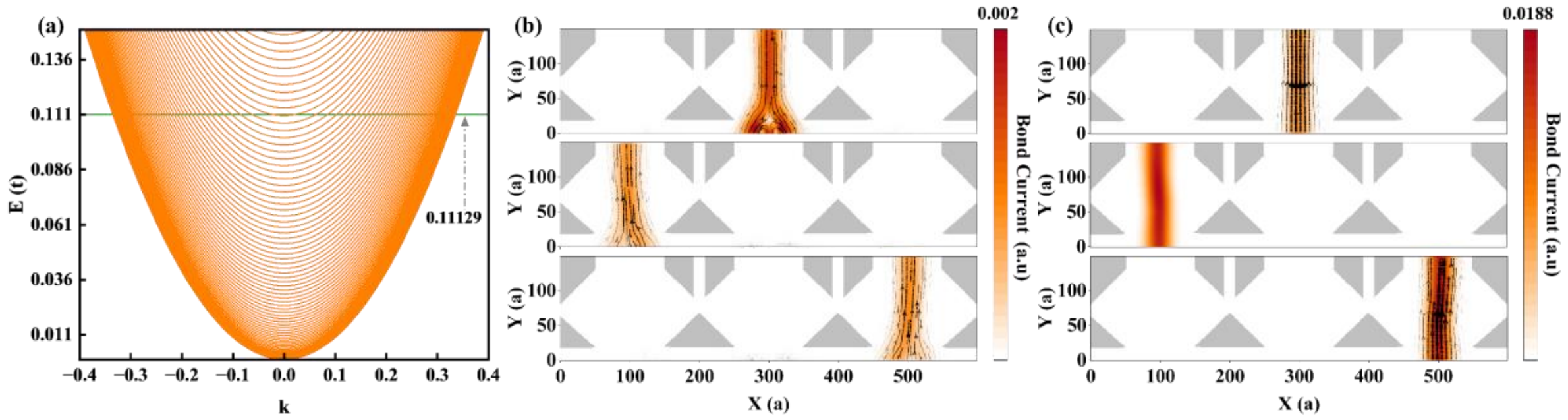

Fig. 4. (a) Dispersion relation of the source lead showing the propagating modes available at the selected electron energy. (b) Bond-current distribution corresponding to a single propagating lead mode. (c) Total bond-current distribution obtained by summing contributions from all propagating lead modes at resonance. The resulting transport remains strongly localized to the resonant output channel.

To further verify the transport pathways, we calculate the bond-current density for the resonant modes $R_{11}$, $R_{21}$, and $R_{31}$, corresponding to the first excited states of the dots. The lead-mode dispersion relation for the input lead is shown in Fig. 4(a). At the selected electron energy, multiple propagating lead modes contribute to transport, as indicated by the number of dispersion branches intersecting the chosen energy line. Each propagating lead mode contributes to the scattering state and therefore to the total bond-current density[27]. Figure 4(b) shows the bond current density corresponding to the first propagating lead mode, while Fig. 4(c) presents the total bond current density obtained by summing the contributions from all propagating lead modes[28]. The resulting bond-current maps reveal highly localized transport pathways only between the source and the resonant output channel, with negligible current leakage to the remaining outputs. The selectivity, calculated as per Equation (4), approaches unity for all resonance peaks, as summarized in Table I. We define it as a ratio of the conductance $G_{SD_i}$ (source, $S$ to drain $D_i$) and the total conductance of all three routes. These results confirm the strong channel selectivity and controllable resonant transport behavior of the proposed multi-dot architecture.

$$\text{Selectivity} = \frac{G_{SD_i}}{\sum_i G_{SD_i}} \tag{4}$$

**TABLE I. Selectivity strength for each of the resonance peaks**

| Resonance Peak | Dominating Dot | Selectivity |
|---|---|---|
| $\mathbf{R_{10}}$ | Left | 0.999990 |
| $\mathbf{R_{11}}$ | Left | 0.999875 |
| $\mathbf{R_{20}}$ | Center | 0.999961 |
| $\mathbf{R_{21}}$ | Center | 0.999670 |
| $\mathbf{R_{30}}$ | Right | 0.999999 |
| $\mathbf{R_{31}}$ | Right | 0.999977 |

Although the present study employs a single-particle model and considers the energy spacing between discrete states in each dot, in practice, this energy will be supplemented by the charging energy, such that $E_{add} = \Delta E + e^2/C_\Sigma$, where $C_\Sigma$ is the total capacitance of the quantum dot. Since this additional charging term only adds to the orbital spacing, Coulomb blockade resonances are expected to be spaced even further apart than the orbital resonances calculated here, thereby further improving channel separation and reducing overlap between neighboring transport windows. The same operating principle is expected to remain valid in multielectron quantum-dot devices as well.

The energy spacing can be increased still further through stronger electrostatic confinement, for example, by making the barrier-gate voltage more negative or by reducing the effective dot size. The present simulations also assume zero electron temperature; in realistic dilution-refrigerator-based experiments, finite electron temperature will primarily broaden the resonance peaks. Provided this thermal broadening remains small compared with the resonance spacing, which is a reasonable assumption, the overall routing behavior and channel selectivity are expected to be largely unaffected. We further compare the feasibility of our simulation with realistic parameter estimation for a Si/SiGe-based heterostructure, as reported in various references[29–34] (See Supplementary Material S2). At an operating temperature of 10 mK, the thermal energy $k_B T$ is approximately $0.86\,\mu eV$, which is substantially smaller than the calculated minimum orbital level spacing $0.89\ meV$ (See Supplementary Material S3). This energy difference is similar to that reported in various studies[35,36]. Thus, thermal broadening is expected to be negligible compared with the energy separation between adjacent orbital states. At finite temperature, however, thermal broadening may become relevant when the thermal energy approaches the resonance linewidth or the energy separation between neighboring resonances.

In this work, we have proposed and theoretically investigated a multi-channel gate-defined quantum-dot architecture consisting of three parallel quantum dots that share a common source reservoir but couple to independent drain channels. Combining an analytical electrostatic model with tight-binding quantum transport simulations, we showed that well-confined quantum-dot orbitals form in each dot and give rise to highly selective resonant electron transport through individual channels. Closed-system eigenstate analysis confirmed the presence of localized orbital states, while open-system transport simulations revealed distinct resonant conductance peaks corresponding to the alignment of these discrete dot levels with the incoming electron energy; the excellent agreement between the transport resonances and the calculated eigenvalue evolution validates the underlying resonant-tunneling mechanism. Careful engineering of the gate geometry and capacitive coupling strengths produced clear separation between neighboring resonances, minimizing channel overlap and enhancing selectivity, while bond current density simulations confirmed spatially selective current flow with negligible leakage into non-resonant channels, yielding selectivity values approaching unity. Because the architecture is scalable to a larger number of channels through suitable gate optimization and is compatible with established 2DEG platforms such as Si/SiGe and GaAs/AlGaAs, it is readily realizable with current nanofabrication technology. The proposed device thus offers a promising route toward programmable quantum routing, selective electron transport networks, and future quantum electronic circuitry based on controllable resonant transport phenomena.

**Acknowledgements:**
We acknowledge the National Quantum Mission, an initiative of the Department of Science and Technology, Government of India, for funding and Rishabh Bharadwaj for providing computing resources.

Supplementary Material

# A Resonant-Tunneling Quantum-Dot Demultiplexer for Scalable Cryogenic Single-Electron Routing

Rajat Bharadwaj[1], Harshita Bose[2], and Madhu Thalakulam[1*]

[1]Indian Institute of Science Education & Research Thiruvananthapuram, Kerala 695551, India

[2]Delhi Technological University, New Delhi 110042, India

## S1. Generalized dimensions used for the model simulation

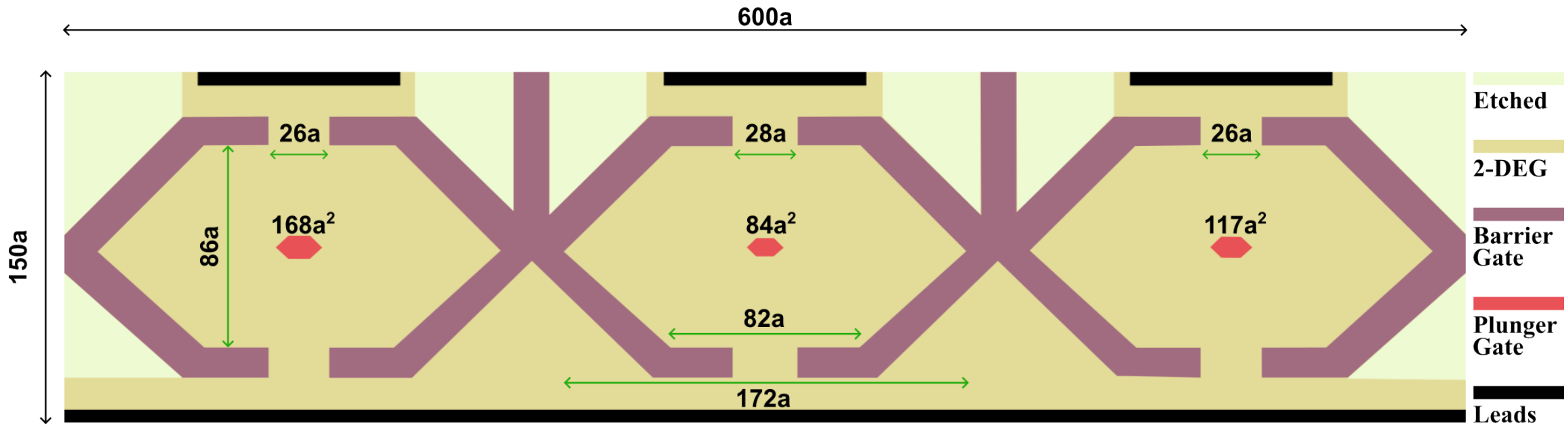

Generalized Dimensions used for the model simulation with respect to the lattice parameter a. Plunger-gate dimensions are given as areas.

## S2. Realistic parameter realization

Throughout the analysis, we use the computational grid spacing $a$and nearest-neighbor hopping energy $t$ to define the device dimensions and corresponding energy scales. For a realistic estimate, we take the example of a Si/SiGe heterostructure.

### Parameters selection

We assume the grid spacing $a = 1\ nm$ and effective electron mass $m_e^* = 0.19m_e$. For the gate voltages to the on-site energy calculation, we assume a reasonable lever arm $\alpha = 0.10\ eV/V$.

* madhu@iisertvm.ac.in

**Representative material and device parameters used to map the dimensionless simulation to a Si/SiGe heterostructure**

| Parameter | Realistic baseline value |
|---|---|
| Lattice Size | $600\ nm \times 150\ nm$ |
| SiGe barrier | $Si_{0.7}Ge_{0.3}$ |
| 2DEG depth | $50\ nm$ |
| Relative dielectric constant | $\epsilon \approx 12$ |
| Effective electron mass | $m^* = 0.19m_e$ |

## S3. Energy-level spacing

Based on the parameters chosen in S2, we calculate the derived quantity $t = 200.5\ meV$. The maximum gate sweep for realizing resonances is performed up to $-1.4t$, as per manuscript Fig. 3(b), which would correspond to $V_{gateActualExtreme} = (-1.4 \times 200.5\ meV)/\alpha = -2.807\ V$. The resonance values can be scaled using this reference, which typically falls within the typical gate-voltage range for these devices.

**Orbital level spacings between the ground and first excited states of the three quantum dots obtained from the closed-system eigenstates.**

| Energy levels | Associated Dot Level Spacing |
|---|---|
| $E_{10}$ $and$ $E_{11}$ | $0.89\ meV$ |
| $E_{20}$ $and$ $E_{21}$ | $0.94\ meV$ |
| $E_{30}$ $and$ $E_{31}$ | $0.89\ meV$ |